\documentclass[8pt,a4paper]{article}
\usepackage[utf8]{inputenc}
\usepackage{amsmath}
\usepackage{amsfonts}
\usepackage{amssymb}
\usepackage{graphicx}
\usepackage{xcolor}
\usepackage{url}
\usepackage{braket}
\usepackage{tikz}
\usepackage[left=3cm, right=3cm, top=2cm, bottom=2cm]{geometry}
\usepackage{rotating}
\usepackage{lscape}
\usepackage{hyperref}
\usepackage{lscape}
\usepackage{wrapfig}
\usepackage{float}

\DeclareMathOperator{\sgn}{sgn}
\DeclareMathOperator{\tr}{tr}
\newcommand{\rl}[1]{\left(#1\right)}

\allowdisplaybreaks

\def\ad{a^\dagger}

\def\H{{\cal H}}

\def\ad{a^\dagger}

\def\be{\begin{equation}}
\def\ee{\end{equation}}
\def\ba{\begin{array}{l}}
	\def\ea{\end{array}}
\def\bea{\begin{eqnarray}}
\def\eea{\end{eqnarray}}
\def\beas{\begin{eqnarray*}}
	\def\eeas{\end{eqnarray*}}

\def\ket#1{| #1 \rangle}
\def\bra#1{ \langle #1 |}

\begin{document}

	\baselineskip 24pt
	
	\begin{center}
		
		{\large \textbf{QUANTUM ENTANGLEMENT OF ANYON COMPOSITES}\par}
		
	\end{center}
	
	\vskip .5cm
	\medskip
	
	\vspace*{4.0ex}
	
	\baselineskip=18pt
	
	\centerline{\large\rm  N Ramadas, V V Sreedhar}
	
	\vspace*{4.0ex}

	\centerline{\it Chennai Mathematical Institute,  SIPCOT IT Park, Siruseri, Chennai, 603103 India} 
	
	\vspace*{1.0ex}

	\vspace*{1.0ex}
	\centerline{ \href{mailto:ramadasn@cmi.ac.in}{\texttt{ramadasn@cmi.ac.in}}, \href{mailto:sreedhar@cmi.ac.in}{\texttt{sreedhar@cmi.ac.in}} }

	\vspace*{5.0ex}

	\begin{abstract}
Studying quantum entanglement in systems of indistinguishable particles, in 
particular anyons, poses subtle challenges. Here, we investigate a model of 
one-dimensional anyons defined by a generalized algebra. This algebra has the
special property that fermions in this model are composites of anyons. A 
Hubbard-like Hamiltonian is considered that allows hopping between nearest 
neighbour sites not just for the fundamental anyons, but for the fermionic 
anyon composites. Some interesting results regarding the quantum entanglement 
of these particles are obtained. 
	\end{abstract}

A fundamental or elementary particle is, by definition, a particle that is not
a composite of other particles. An interesting question concerns whether, and 
when, a given particle should be considered truly elementary. The answer to 
this question depends on the scale of the energy that is relevant to the system 
we are interested in. In the standard model of particle physics, which is the 
best known theoretical fit for all the experimental results we have so far, 
at the highest energies accessible in the laboratories, the fundamental 
particles are quarks, leptons, and gauge bosons. It has been postulated that 
these `fundamental' particles are composed of even smaller particles called 
preons \cite{raitioModelLeptonQuark1980}, or perhaps extended objects like 
strings \cite{polchinski1998string}, but there is no experimental evidence to 
support these ideas at the moment. 

On the other hand, composites of fundamental particles themselves behave like
individual particles and exhibit collective behaviour which is much more 
complex than that of the constituent particles. In the BCS theory, for 
example, the superconducting transition is described by the condensation of 
Cooper pairs \cite{cooperBoundElectronPairs1956} which, as the name suggests,
are made up of pairs of fermions. There are theories of composite bosons 
(cobosons) \cite{combescotManybodyPhysicsComposite2008} that are composites of 
an even number of fermions and they behave like bosons under certain 
conditions. Several bulk phenomena in condensed matter physics are understood 
in terms of quasiparticles and collective excitations like phonons, magnons, 
spinons, plasmons, and polarons \cite{kittel1987quantum}. 

We mention the example of fractional quantum Hall effect 
\cite{girvin1987quantum} which is of particular interest for the purposes of 
this paper. One way to understand this system comes from the composite fermion 
picture \cite{jainCompositefermionApproachFractional1989}. It should be noted 
that the said composite fermion is actually an even number of fluxes attached 
to an elementary electron. 

Another interesting composite particle which has captured the imagination of 
physicists, with applications not only to the fractional quantum Hall effect 
but also to topological quantum computation, was proposed by Wilczek 
\cite{wilczek1990fractional} and is called an anyon. It is a composite of a 
charged particle and a magnetic flux tube. Anyons obey the so-called exchange 
statistics --  a generalization of Bose-Einstein and Fermi-Dirac statistics 
associated with the transformation properties of  multi-particle wave 
functions of indistinguishable particles when an exchange of two particles 
is performed. Wilczek's model may be considered as a physical realization of 
the ideas of Leinaas and Myrheim \cite{leinaasTheoryIdenticalParticles1977} 
who had argued earlier that the exchange statistics is a consequence of the 
nontrivial topology of the classical configuration space of indistinguishable 
particles in low dimensions. 

Both Jain's and Wilczek's constructions are reductionist in nature, defining 
an anyon in terms of more elementary objects like charged particles and 
fluxes. It is also possible to define anyons more holistically, as objects 
which are fundamental or elementary by definition -- a task accomplished by
suitably generalising their quantum operator algebra and Pauli's exclusion 
principle. In doing so, one is not restricted to low dimensions. Some of 
these approaches include Gentile statistics \cite{gentile1940itosservazioni}, 
Green's statistics \cite{greenGeneralizedMethodField1953}, quons 
\cite{greenbergExampleInfiniteStatistics1990}, and Haldane's exclusion 
statistics \cite{haldaneFractionalStatisticsArbitrary1991}.  

In the present work, we propose one such model of anyons by defining a 
generalized algebra. Since anyons are fundamental particles in this model,
there is no need to introduce fluxes. A special feature of this model is 
that we can construct fermions as composites of anyons, which is exactly 
the opposite in spirit to Jain's and Wilczek's constructions.

As an application of this model, we construct a tight-binding Hamiltonian 
of particles hopping on a one-dimensional lattice, allowing not just the 
elementary anyons to hop, but also the fermionic anyon-composites to hop from 
one site to the other. We solve the two-site model explicitly for two special 
cases. Our primary motivation for introducing the above model is to investigate
the quantum entanglement properties of anyons, in particular the dependence on 
the statistics parameter. In earlier papers, we have used the information 
theoretic methods developed by Lo Franco and Compagno in the context of 
entanglement of fermions and bosons, to study the entanglement in the 
Leinaas-Myrheim model and the two-site anyonic Hubbard model. The model 
analyzed in this paper is a natural, but significant, generalization of the 
latter model. The interplay between anyons and fermions leads to interesting 
features in the dependence of the entanglement entropy on the statistical 
parameter.

The rest of the paper is organised as follows: In section 1, we review the 
information theoretic approach to entanglement of indistinguishable particles.
In section 2, we introduce our model of anyons and discuss the construction 
of fermions as composites of anyons. In section 3, we introduce the 
tight-binding model of anyons and their composites, and derive some interesting 
results regarding quantum entanglement. In section 4, we present the 
conclusions. Some calculational details are relegated to the Appendices. 

\section{Information theoretic approach to quantum entanglement}
\label{s:Information theoretic approach}
Standard measures to quantify entanglement of distinguishable particles 
like the Schmidt rank and the von Neumann entropy fail to work in the case of 
indistinguishable particles. These measures give the wrong result that
the state is entangled, even if the only correlations present in the state 
are due to the indistinguishability of particles. 

Several alternate approaches \cite{franco2016quantum,balachandran2013algebraic,
ghirardiIdenticalParticlesEntanglement2005a,
schliemannQuantumCorrelationsTwofermion2001a,
paskauskasQuantumCorrelationsTwoboson2001a,wiseman2003entanglement,
killoran2014extracting,franco2018indistinguishability,benatti2014entanglement}
were developed to overcome the issues related to the quantum entanglement of 
indistinguishable particles. These approaches propose new notions of 
indistinguishable particle entanglement and measures to quantify it. In this 
section, we discuss the information theoretic approach as developed by Lo 
Franco and Compagno \cite{franco2016quantum}, which allows us to define the 
notions of partial trace and reduced density matrix in the case of 
indistinguishable particles. The von Neumann entropy computed using the 
reduced density matrix thus defined is proposed as a measure to quantify 
the entanglement of indistinguishable particles.

In the standard approach in quantum mechanics, indistinguishable particles 
are assigned labels and the states of bosons (fermions) are obtained by 
symmetrizing (antisymmetrizing) with respect to these labels. But, these 
labels are unphysical since indistinguishable particles cannot be individually 
addressed. Also, these labels add a fictitious contribution to the real 
entanglement of the particles.

The information theoretic approach to quantum entanglement of 
indistinguishable particles does not involve labeling individual particles. 
The state of indistinguishable particles is considered a holistic indivisible 
entity. For example, a state of a system of two indistinguishable particles is 
simply represented as $ \ket{\psi,\phi} $ where $ \psi $ and $ \phi $ describe 
single particle states. The relevant quantities in the approach are transition 
amplitudes which can be expressed in terms of the single-particle transition 
amplitudes. For example, if we consider the transition amplitude from the 
two-particle state $ \ket{\psi,\phi} $ to the state $ \ket{\zeta,\varphi} $, 
it can be expressed in terms of the single-particle transition amplitudes in 
the following way
\begin{align}
\begin{aligned}
\braket{ \psi,\phi |\zeta,\varphi } & = \braket{\psi|\zeta} 
\braket{\phi|\varphi} + \eta \braket{\psi|\zeta} \braket{\phi|\varphi} 
\end{aligned}
\end{align}
where $ \eta=1 $ for bosons and $ \eta=-1 $ for fermions.
It is also possible to define an inner product between Hilbert spaces of 
different dimensionality. For example, the inner product between a 
two-particle state $ \ket{\varphi_1,\varphi_2} $ and single-particle 
state $ \ket{\psi} $ is defined as follows
\begin{align}
\begin{aligned}
\braket{\psi|\cdot |\varphi_1,\varphi_2} := \braket{\psi| \varphi_1,\varphi_2} 
= \braket{\psi|\varphi_1} \ket{\varphi_2}+\eta \braket{\psi|\varphi_2}
\ket{\varphi_1}
\end{aligned}
\end{align}
where $ \eta = \pm 1 $. The above formalism can be generalized to the 
$N- $particle case. 

The definition of the inner product between Hilbert spaces with different 
dimensions allows us to define a partial trace operation and the reduced 
density matrix. We explain the construction of the one-particle reduced density 
matrix corresponding to an $N-$particle state $ \ket{\Phi}$: Consider a 
basis $\lbrace\ket{k} \rbrace $ of the single-particle Hilbert space. The 
normalized $(N-1)$-particle state obtained by projecting $\ket{\Phi} $ onto 
$ \ket{k} $ is given by
\begin{align}
\begin{aligned}
\ket{\phi_k} = \frac{\braket{k|\Phi}}{\sqrt{\braket{\Phi|\Pi^{(1)}_k|\Phi}}}
\end{aligned}
\end{align}
where $ \Pi^{(1)}_k = \ket{k} \bra{k} $. The probability to get a state
$\ket {k} $ after projection is
\begin{align}
p_k = \frac{1}{N}\braket{\Phi|\Pi^{(1)}_k|\Phi}
\end{align}
Then the reduced density matrix is defined as
\begin{align}\label{infreduced}
\rho^{(N-1)} = \sum_k p_k \ket{\phi_k} \bra{\phi_k}
\end{align}
The entanglement entropy is found by computing the von Neumann entropy of 
the reduced density matrix using the standard formula
\begin{align}\label{entropy}
E[\Phi] = S[\rho^{(N-1)}]= -\text{Tr} \rl{\rho^{(N-1)} \ln \rho^{(N-1)}} 
= -\sum_i \lambda_i \ln \lambda_i
\end{align}
where $\lambda_i$ is an eigenvalue of the reduced density matrix. The 
entanglement entropy defined above allows us to quantify the entanglement 
in a state of indistinguishable particles.

The method can be recast in the language of second quantization which is 
useful to generalize the approach to study anyons. We note that the state 
obtained by taking the inner product of a single particle state $ \ket{k} $ 
with an $ N- $particle state $ \ket{\Phi} $ is equivalent to the state 
obtained by the action of the annihilation operator $ a_k $ corresponding 
to the state $ \ket{k} $ on the state $ \ket{\Phi} $, that is, 
\begin{align}
\begin{aligned}
a_{k} \ket{\Phi} \equiv \bra{k} \cdot \ket{\Phi}
\end{aligned}
\end{align}
where $ a_{k}$ is the annihilation operator such that 
$ a^\dagger_{k} \ket{0} = \ket{k}$ and the right hand side denotes the 
inner product between a single particle state $ \ket{k} $ and the 
$N-$particle state $\ket{\Phi}$. Using the above equivalence, the formula for 
the one-particle reduced density matrix in the second quantization language is
\begin{align}\label{infred}
\rho^{(N-1)} \rl{\Phi}  = \frac{1}{\mathcal{N}}\sum_k a_{k}\ket{\Phi}\bra{\Phi} 
a^\dagger_{k} 
\end{align}
where $\mathcal{N}= \braket{\Phi|\sum_k a^\dagger_{k}a_{k}|\Phi} $. Note that 
the above expression allows us to compute the reduced matrix in the case of 
anyons by using the creation and annihilation operators of anyons for a given 
initial multi-anyon state. After obtaining the reduced density matrix, the 
entanglement entropy is computed using the expression in Eq. \ref{entropy}. 

\section{A model for anyons}
There are many approaches to generalizing Bose-Einstein and Fermi-Dirac 
statistics \cite{gentile1940itosservazioni,greenGeneralizedMethodField1953,
greenbergExampleInfiniteStatistics1990,
haldaneFractionalStatisticsArbitrary1991}. One approach is based on the change 
in the properties of multiparticle wave functions of a system of 
indistinguishable particles when any two particles are exchanged. In the case 
of bosons (fermions), the wave function picks up a factor equal to $\pm 1$ 
when any two particles are exchanged. In the case of anyons, the 
wavefunction can pick up a phase factor $ e^{i\theta} $ where $ \theta \in 
[0,\pi]$ can have fractional values. Another approach to generalizing 
Bose-Einstein and Fermi-Dirac statistics is based on generalizing Pauli's 
exclusion principle. 

In the present work, we propose anyons in one dimension which are constructed 
by generalizing the phase factor picked up by the multiparticle wave function 
under exchange of particles, and by generalizing the number of particles 
allowed to occupy a quantum state. The latter case corresponds to Gentile 
statistics \cite{gentile1940itosservazioni}.

Consider a system of anyons in one dimension. Let $ a^\dagger  (a)$ denote 
the creation (annihilation) operators of anyons. We propose the following 
commutation relations satisfied by the creation and annihilation operators
\begin{align} \label{eq:algebra}
\begin{aligned} 
a_j a_k & =e^{i \theta \sgn(j-k)  } a_k a_j,~~~~~ j \neq k\\
a_j a_k^\dagger & =e^{-i \theta \sgn(j-k)  } a_k^\dagger a_j ~~~~ j\neq k\\
a_j a_j^\dagger & = 1-  (a_j^\dagger)^{(\nu)} (a_j)^{(\nu)} \\
(a_j^\dagger)^{\nu+1}&=0
\end{aligned}
\end{align}
Here, $ j $ and $ k $ denote modes or sites, $ \theta \in [0,\pi] $ and 
$ \nu \in \mathbb{N} $, the set of natural numbers. The function $ \sgn(x) $ 
denotes the sign function where $ \sgn(x)=1 $ if $ x>0 $, $ \sgn(x)=0 $ if 
$ x=0 $ and $ \sgn(x)=-1 $ if $ x<0 $.

The first two relations in equation \ref{eq:algebra} indicate that the 
multi-particle wave function of anyons picks up a non-trivial phase factor 
when two particles are exchanged. The last relation $(a_j^\dagger)^{\nu+1}=0$ 
implies that the maximum number of particles allowed in a given node or site 
is $\nu$. This corresponds to a generalization of Pauli's exclusion principle. 
Restriction in the occupancy of particles in a given quantum state to an 
integer $ \nu $ gives rise to particles obeying Gentile statistics 
\cite{gentile1940itosservazioni}. We will show that the distribution function 
of the anyons considered here is the same as the Gentile's distribution 
function in section \ref{distribution}.

In the limit $ \theta= \pi $ and $ \nu=1 $, Eq. \ref{eq:algebra} reduces to 
anticommutation relations which represent fermions. Psuedo-bosons are 
obtained in the limit $ \theta=0 $ and $ \nu= 0 $ where the particles 
behave as bosons when two particles are exchanged, but the occupancy is 
restricted to one particle per site or mode. 

The limit $ \nu=\infty $ and $ \theta=0 $ indicate that the wave function 
is unchanged when any two particles are exchanged and any number of 
particles are allowed to occupy the same site or mode. These are the 
characteristics of bosons. But, in this limit, the algebra in \ref{eq:algebra} 
does not give the usual commutation relations. Nevertheless, by looking at the
properties of the particles, we can say that these particles are bosons in 
this limit. 

For general values of $ \theta $ and $ \nu $, the algebra represents anyons. 

The total number operator for anyons is defined as 
\begin{align}
\begin{aligned}
\hat{N} = \sum_j{n_j},~~~n_j = \sum_{k=1}^{\nu} (\ad_j)^k a_j^k 
\end{aligned}
\end{align}
It satisfies the standard algebraic relations
\begin{align}
\begin{aligned}
\left[\hat{N},a_j\right] = -  a_j,~~~~\left[\hat{N},a_j^\dagger\right] =  
a_j^\dagger
\end{aligned}
\end{align}
A derivation of these results can be found in appendix \ref{numberoperator}.  

We define the Fock vacuum $ \ket{0} $ such that $ a_j \ket{0}=0 $ for all 
$ j $ values. The states of anyons are constructed by acting with creation 
operators on the vacuum state. This completely specifies the Fock space 
$ \mathcal{F}_a $ of our anyons.
\subsection{Distribution function}\label{distribution}
We consider a free gas of anyons in the grand canonical ensemble with the 
following Hamiltonian
\begin{align}
\begin{aligned}
H_{GCE}=\sum_j \epsilon_j n_j
\end{aligned}
\end{align}
where $ n_j $ is the number density operator. The partition function is given 
by $ Z = \tr (e^{ -\beta (H-\mu N)}) $, where $ \beta $ is the inverse 
temperature, $ \mu $ is the chemical potential and $ N $ is the total number 
operator. The distribution function is given by
\begin{align}
\begin{aligned}
\braket{n_j}& =\frac{1}{  \rl{e^{\beta  (\epsilon_j-\mu )} -1}} - 
\frac{ (\nu +1) \rl{e^{\beta (\epsilon_j-\mu )} -1}}
{\rl{e^{\beta {(\nu+1) }(\epsilon_j-\mu )} -1}   }
\end{aligned}
\end{align}
The details of the derivation are given in appendix \ref{app:distri}. Note 
that this is the same as the distribution function of particles obeying 
Gentile statistics \cite{gentile1940itosservazioni}, and the algebra given 
in equation \ref{eq:algebra} represents Gentile particles. However, we mention 
that the parameter $ \theta $ makes statistics of these particles more general 
than Gentile statistics.
\subsection{Fermions as composites of anyons}
As mentioned earlier, the algebra in equation \ref{eq:algebra} represents 
fermions when $ \nu=1 $ and $ \theta=\pi $. But, our interest is to construct 
fermions as composites of anyons -- particles  represented by the algebra in 
equation \ref{eq:algebra} for general values of $ \nu $ and $ \theta $.

Let $ f_j^\dagger $ and $ f_j $ represent fermionic creation and annihilation 
operators respectively. A crucial property of fermions is that they obey 
Pauli's exclusion principle.  This translates into an operator condition 
$ (f_j)^2=0 $ for all $ j $ values. In the algebra in equation 
\ref{eq:algebra}, we note that $ (a_j)^{\nu+1}=0 $. A naive guess {\it viz.} 
defining $ f_j = (a_j)^{\frac{\nu+1}{2}} $ will ensure $f_j^2=0 $. To 
avoid difficulties with fractional powers, we restrict the values of $\nu$ 
to odd integers. Also, we assume that the value of $ \nu $ is finite. For 
convenience, we use the notation $ m= \frac{\nu+1}{2}$, so that $f_j = a_j^m$.

We define the fermionic operators as follows
\begin{align}
\begin{aligned}
f_j = a_j^{m}, ~f^\dagger_j = (a^\dagger_j)^{m},~~~ m= \frac{\nu+1}{2},~~~
\nu =1,3,5,.. 
\end{aligned}
\end{align}
We note the following algebraic relation for distinct values of $ j $ and $ k $ 
\begin{align}
\begin{aligned}
f_j f_k & = e^{i  m^2\theta\sgn(j-k)}  f_k f_j,~j\neq k
\end{aligned}
\end{align}
We require $ f_j $ and $ f_k $ to anticommute for all values of $ j $ and  
$ k $ with $ j\neq k $. Therefore the phase factor 
$  e^{i  m^2\theta\sgn(j-k)} $ should be equal to $ -1 $ for all values of 
$ j,k $ and $ \nu $. This restricts the values of $ \theta $. For our purposes, 
we set 
\begin{equation}
\theta =  \frac{\pi}{m^2} = \frac{4\pi}{ (\nu+1)^2}
\end{equation}
With these choices for $ \theta $ and $ \nu $, it is straightforward to show 
that the operators $ f_j $ and $ f_j^\dagger $ satisfy the anticommutation 
relations
\begin{align}
\begin{aligned}
f_j f_k &= -f_k f_j ,~~ j\neq k\\
f_j f^\dagger_k &= -f^\dagger_k f_j ,~~ j\neq k \\
f_j f_j^\dagger &= 1- f_j^\dagger f_j \\
f_j^2&=0
\end{aligned}
\end{align}
The details of the derivation are given in the appendix \ref{app:fermalg}.

The physical meaning of the construction is that a composite of $ m $ 
anyons at a given site or mode behaves like a fermion. Note that our 
construction of fermions as composites of anyons is different from the 
composite fermions proposed by Jain which involve attaching 
an even number of fluxes to an elementary electron. Jain's composite fermions 
are more like anyons than fermions. Our fermions are composites of anyons. 
\section{A Hubbard-like model}
Migration plays a ubiquitous role in nature. Insects, birds, fish, and animals
migrate in search of food, to breed, and to escape harsh climate. Human beings
migrate to greener pastures, or to escape persecution due to socio-political
reasons, often individually, sometimes in groups. 

In physics, itinerant electrons play an important role in understanding 
magnetism. The Hubbard model \cite{hubbardElectronCorrelationsNarrow1963} 
describes the hopping of electrons from one lattice site to another, along 
with on-site interaction terms, and is useful in understanding metal-insulator 
transitions. In the tight-binding approximation, one considers the cost of 
hopping alone, not the on-site interactions. Only individual particles, not 
their composites, are involved in the hopping.

In an earlier paper \cite{ramadas2022quantum}, we considered the anyonic 
Hubbard model to explore quantum entanglement of anyons. In this paper, we 
consider a scenario where not just the anyons, but also their fermionic 
composites, can hop on a one-dimensional lattice with open boundary 
conditions. We propose the following Hamiltonian for the model 
\begin{align}
\begin{aligned}
H&=H_a+ H_f \\
H_a &= -\kappa_a \sum_{j=1}^{L}\rl{ a_{j+1}^\dagger a_j+a_{j}^\dagger a_{j+1}}
\\
H_f& = -\kappa_f \sum_{j=1}^L\rl{ f_{j+1}^\dagger f_j+f_{j}^\dagger f_{j+1}} 
\end{aligned}
\end{align}
Here $L$ is the number of lattice sites. The term $H_a$ describes the hopping 
of anyons and the term $ H_f $ describes the hopping of fermions. The anyonic 
and fermionic hopping parameters are $ \kappa_a $ and $ \kappa_f $, 
respectively. 

We only consider the nearest-neighbor hopping of particles in this model. 
At first sight, this looks like a double tight-binding model of anyons and 
fermions, where the two have nothing to do with each other. However, if we 
rewrite $ H_f $ in terms of anyonic operators using $ f_j =a_j^m $,
\begin{align}
\begin{aligned}
H_f& =  -\kappa_f \sum_{j=1}^{L} \rl{ (a_{j+1}^\dagger)^m
 (a_j)^m+(a_{j}^\dagger)^m (a_{j+1})^m} 
\end{aligned}
\end{align}
we see that $ H_f $ actually describes nearest-neighbor interaction 
between anyons. The composite nature of fermions is linked to the 
interaction of anyons.

\subsection{Two-site case  }
In this discussion, we consider the simplest case of two lattice sites.  The 
Hamiltonian for the two-site case is given by
\begin{align}
\begin{aligned}
H&=H_a+H_f\\ H_a& = -\kappa_a \rl{ a_{1}^\dagger a_2+a_{2}^\dagger a_{1}} \\
H_f &=-\kappa_f \rl{ (a_{1}^\dagger)^m (a_2)^m+(a_{2}^\dagger)^m (a_{1})^m}
\end{aligned}
\end{align}
Since the total number operator $ \hat{N} $ commutes with the Hamiltonian, the 
total number of anyons is a conserved quantity. The Fock space $\mathcal{F}_a$ 
can be organized into $ N-$anyon Hilbert spaces corresponding to the 
eigenvalues of the total number operator. Since the maximum occupancy at each 
lattice site is $ \nu $, $N$ can take values $ N=0,1,2,..,2\nu $ for a given 
value of $\nu$. The dimension of the $ N- $particle Hilbert space 
$ \mathcal{H}_N $ for a specific value $ \nu $ is $ d(N,\nu) $ where
\begin{align}
\begin{aligned}
d(N,\nu) = \begin{cases}
N+1 ~&\text{if}~0 \leq N\leq \nu \\
2\nu+1-N~&\text{if}~\nu<N\leq 2\nu
\end{cases}
\end{aligned}
\end{align} 
This also gives the dimension of the Fock space in the case of the two-site 
model which is $ \nu^2 $ for a given $ \nu $. To express the Hamiltonian in 
matrix form, we choose the following basis of the Fock space
\begin{align}
\begin{aligned}
\lbrace  \ket{r,s} \rbrace ,~~~r ,s =0,1,2,..,\nu
\end{aligned}
\end{align}
where $\ket{r,s} = (a^\dagger_1)^r (a^\dagger_2)^s\ket{0}$. We use the 
notation that $ X^0 =1 $, for any operator $ X $. The basis of $N-$particle 
Hilbert space is the set of all vectors $\lbrace\ket{r,s}\rbrace$ such that 
$ r+s=N $.

The Hamiltonian can be expressed in the matrix form in the above basis
\begin{align}
\begin{aligned}
\pmb{H}_{q,p;r,s} = \bra{q,p} H \ket{r,s}=& -\kappa_a \rl{e^{\frac{i \pi (1-q)}
{m^2} } \delta_{q,r+1} \delta_{p+1,s}+  e^{\frac{i \pi (r-1)}{m^2}  } 
 \delta_{q+1,r} \delta_{p,s+1} }\\& -\kappa_f \rl{ e^{\frac{i \pi m(1-q)}{m^2}}
\delta_{q,r+m} \delta_{p+m,s}+e^{\frac{i \pi m (r-1)}{m^2}}\delta_{q+m,r} 
\delta_{p,s+m} } 
\end{aligned}
\end{align}
The details of the derivation are given in appendix \ref{app:hammatrix}.
The matrix $ \pmb{H} $ can be written as a direct sum of $ 2\nu $ matrices, 
that is, $\pmb{H} = \sum_{\oplus} \pmb{H}_{N}$ where $ N=0,1,..,2\nu $ and 
$\pmb{H}_{N}  $ is the $N-$particle Hamiltonian matrix. This allows us to 
simplify the problem.

We observe that the matrix $\pmb{H}_{N}$ is unitarily equivalent to a 
Hermitian Toeplitz matrix via a unitary transformation by the diagonal matrix 
with elements $U_{q,p;r,s} = e^{\frac{\pi i}{2m^2} r(r-1) } \delta_{q,r} 
\delta_{p,s} $. The transformed Hamiltonian is given by
\begin{align}\label{toeplitz}
\begin{aligned}
\rl{\pmb{ H}'_N}_{q,p;r,s}
&= -\kappa_a \rl{\delta_{q,r+1} \delta_{p+1,s}+  \delta_{q+1,r} 
\delta_{p,s+1}} -\kappa_f \rl{e^{\frac{-i \pi (m-1)}{2m} }  \delta_{q,r+m} 
\delta_{p+m,s}+ e^{\frac{i \pi (m-1)}{2m} }  \delta_{q+m,r} \delta_{p,s+m}}
\end{aligned}
\end{align}
It is evident from the expression that the matrix $ \pmb{ H}'_N $ has zeroes 
everywhere except four diagonals parallel to the main diagonal. The spectra
of these Toeplitz matrices is not known analytically. However, we may look at 
two special cases where the spectrum can be obtained analytically.
 
\subsection{The special case $ \kappa_f = 0 $} 
In the case when $ \kappa_f=0 $, the model corresponds to a 
tight-binding model of anyons represented by the Hamiltonian
\begin{align}
\begin{aligned}
H_a&=-\kappa_a \rl{ a_{1}^\dagger a_2+a_{2}^\dagger a_{1}}
\end{aligned}
\end{align}
From equation \ref{toeplitz}, it is clear that the $ N $- particle Hamiltonian 
in the matrix form is a tridiagonal Hermitian Toeplitz matrix with zero 
entries along the main diagonal. The spectra of these matrices are known 
analytically \cite{noscheseTridiagonalToeplitzMatrices2013}. Using standard 
formulae, the energy eigenvalues of the $ N $-particle 
Hamiltonian are given by
\begin{align}
\begin{aligned}
\epsilon^{(a)}_j & =- 2 \kappa_a \cos \rl{\frac{j \pi}{d(N,\nu)+1}},
~j=1,2,..,d(N,\nu)
\end{aligned}
\end{align}
The components of the eigenvector $\phi_N^{j} = [\xi_{j,1},\xi_{j,2},...,
\xi_{j,N}]^T$ are
\begin{align}
\begin{aligned}
\xi_{j,k} & = e^{\frac{i\pi}{2m^2} k(k-1)}\sin \frac{jk\pi}{d(N,\nu)+1},~j,k
=1,2,..,d(N,\nu)
\end{aligned}
\end{align}
\subsection{The special case $ \kappa_a = 0 $} 
In the case when $ \kappa_a=0 $, the Hamiltonian which is given by
\begin{align}
\begin{aligned}
H_f& = -\kappa_f \rl{ f_{1}^\dagger f_2+f_{2}^\dagger f_{1}} 
\end{aligned}
\end{align}
describes a tight-binding model of fermions. 
The matrix elements of the Hamiltonian are
\begin{align}\label{hfmatrixform}
\begin{aligned}
(H_f)_{q,p;r,s} = \bra{q,p} H \ket{r,s}&=  -\kappa_f \rl{ e^{\frac{i 
\pi m(1-q)}{m^2} }\delta_{q,r+m} \delta_{p+m,s}+e^{\frac{i \pi m (r-1)}
{m^2}  }  \delta_{q+m,r} \delta_{p,s+m} }
\end{aligned}
\end{align}

Note that if $ N<m $, there are not enough numbers of anyons to form a 
fermion. Therefore the $ N $-particle Hamiltonian corresponding to the 
fermionic hopping term is zero if $ N<m $. Also, if $ N>3m-2 $, both sites
are occupied by one fermion each. Since both sites are occupied by one fermion 
each, it is not possible for a fermion to hop from one lattice site to the 
other.  Thus, the $ N $-particle Hamiltonian corresponding to the fermionic 
hopping term is zero if $ N>3m-2 $. Therefore, we only need to consider the 
cases where $ m \leq N \leq 3m-2 $.

Consider the $ N- $particle Hilbert space basis $  \lbrace \ket{\chi^{(N)}_j} 
\rbrace  $ where 
\begin{equation}
\ket{\chi^{(N)}_j} = (a^\dagger_1)^j (a^\dagger_2)^{(N-j)} \ket{0}
\end{equation}
and 
\begin{align*}
\begin{aligned}
j& = \begin{cases}
0,1,..,N ~~&\text{if}~~ 0 \leq N\leq 2m-1 \\
N-(2m-1),N-(2m-1)+1,..,(2m-1) ~~&\text{if}~~ 2m-1 < N\leq 2(2m-1) \\
\end{cases}
\end{aligned}
\end{align*}
We expand the $ N $-particle Hamiltonian $ H^{(N)}_f $ as follows
\begin{align}
\begin{aligned}
H^{(N)}_f &= \sum_{j,k} \braket{\chi^{(N)}_j|H_f|\chi^{(N)}_k }
\ket{\chi^{(N)}_k  } \bra{\chi^{(N)}_j }\\
&=-\kappa_f \sum_{j,k} \rl{ e^{\frac{i\pi m(1-j)}{m^2} }\delta_{j,k+m}
	+e^{\frac{i \pi m (k-1)}{m^2}  }  
\delta_{j+m,k}  } \ket{\chi^{(N)}_k  } \bra{\chi^{(N)}_j } \\
& = -\kappa_f \sum_k \rl{ e^{\frac{i \pi m(1-k-m)}{m^2} } \ket{\chi^{(N)}_k  } 
\bra{\chi^{(N)}_{k+m} }+e^{\frac{i \pi m (k+m-1)}{m^2} } 
\ket{\chi^{(N)}_{k+m}  } \bra{\chi^{(N)}_k }  }
\end{aligned}
\end{align}
Note that the summation limits depend on $ N $
\begin{align*}
\begin{aligned}
k& = \begin{cases}
0,1,..,N ~~&\text{if}~~ 0 \leq N\leq 2m-1 \\
N-(2m-1),N-(2m-1)+1,..,(2m-1) ~~&\text{if}~~ 2m-1 < N\leq 2(2m-1) \\
\end{cases}
\end{aligned}
\end{align*}
First, we consider the case where $ m\leq N\leq 2m-1 $. The Hamiltonian is 
\begin{align}
\begin{aligned}
H^{(N)}_f &= -\kappa_f \sum_{k=0}^{N-m} \rl{ e^{\frac{i \pi m(1-k-m)}{m^2} } 
\ket{\chi^{(N)}_k }\bra{\chi^{(N)}_{k+m}} 
	+e^{\frac{i \pi m (k+m-1)}{m^2}}\ket{\chi^{(N)}_{k+m}  } 
\bra{\chi^{(N)}_k }  }
\end{aligned}
\end{align}
Consider a term $ \ket{\chi^{(N)}_{k'}  } \bra{\chi^{(N)}_{k'+m} }  $ and 
another term $ \ket{\chi^{(N)}_{k''}  } \bra{\chi^{(N)}_{k''+m} }  $ for 
different values $ k' $ and $ k'' $ of the summation index $ k $. We note 
that $ k'' $ cannot be equal to $ k'+m $. If $ k''=k'+m $, the second term 
is $ \ket{\chi^{(N)}_{k'+m}  } \bra{\chi^{(N)}_{k'+2m} }$. Since maximum 
occupancy at a given site is $ \nu = 2m-1 $, the second term vanishes. This 
property allows us to uniquely pair basis states and form orthonormal states 
defined by
\begin{align}
\begin{aligned}
\ket{\xi_{\pm,k}^{(N)}} & =\frac{1}{\sqrt{2}} \rl{ e^{\frac{i \pi m(1-k-m)}
{2m^2} } \ket{\chi^{(N)}_k  } \pm  e^{-\frac{i \pi m(1-k-m)}{2m^2} }   
\ket{\chi^{(N)}_{k+m} } }
\end{aligned}
\end{align}
The Hamiltonian $ H^{(N)}_f  $ is diagonal in these states
\begin{align}
\begin{aligned}
H^{(N)}_f &= -\kappa_f \sum_{k=0}^{N-m} \rl{\ket{\xi_{+,k}^{(N)}} 
\bra{\xi_{+,k}^{(N)}} -  \ket{\xi_{-,k}^{(N)}} \bra{\xi_{-,k}^{(N)}} }
\end{aligned}
\end{align}
Therefore the non-zero eigenvalues are $ \pm \kappa_f $ with corresponding 
eigenstates $ \ket{\xi_{\mp,k}^{(N)}}  $. 

Using a similar method, the eigenvalues and eigenstates of the Hamiltonian 
$ H^{(N)}_f  $ can be obtained for the case where $ 2m-1<N \leq 3m-2 $, by a 
change in the summation limits. The non-zero eigenvalues can be shown to be 
$ \pm \kappa_f $ in this case. 

The Hamiltonian $ H $ is difficult to solve for general cases of $\nu $. 
Hence, we use numerical methods to solve the eigenvalue equation 
\begin{align}\label{eq:eigenvalue}
\begin{aligned}
H_N \ket{\Phi_N^{(j)}} & = \epsilon_N^{(j)}\ket{\Phi_N^{(j)}}
\end{aligned}
\end{align}
where $ \epsilon_N^{(j)}  $ is $ j^{th} $ energy eigenvalue and 
$ \ket{\Phi_N^{(j)}} $ is $ j^{th} $ energy eigenstate of the 
$ N- $particle Hamiltonian.
\section{Entanglement entropy}
In this section, we use the information theoretic approach to study quantum 
entanglement of anyons and their fermionic composites. In particular, we are 
interested in studying how the entanglement entropy depends on the parameter 
$m=(\nu+1)/2$ which represents both anyonic statistics and the composite nature
of fermions.  

From Eq \ref{entropy}, it is clear that the entanglement entropy depends on 
the initial $N-$anyon state $\ket{\Phi} $. Since the properties of the 
$N-$particle Hilbert space depend on the statistics parameter $ \nu =2m-1$ 
and the number of particles $N$, it is natural to expect that the entanglement 
entropy corresponding to an $N$-anyon state depends on $\nu$ and $N$. To study 
the dependence, we find the entropy corresponding to the ground state of the 
$N-$particle Hamiltonian by computing the one-particle reduced density matrix 
for different values of $ \nu $ and $ N $. In the case of ground state 
degeneracy, we randomly choose one of the states. This is justified since 
we aim at a qualitative understanding of the dependence.

\begin{figure}
	\includegraphics[scale=0.5]{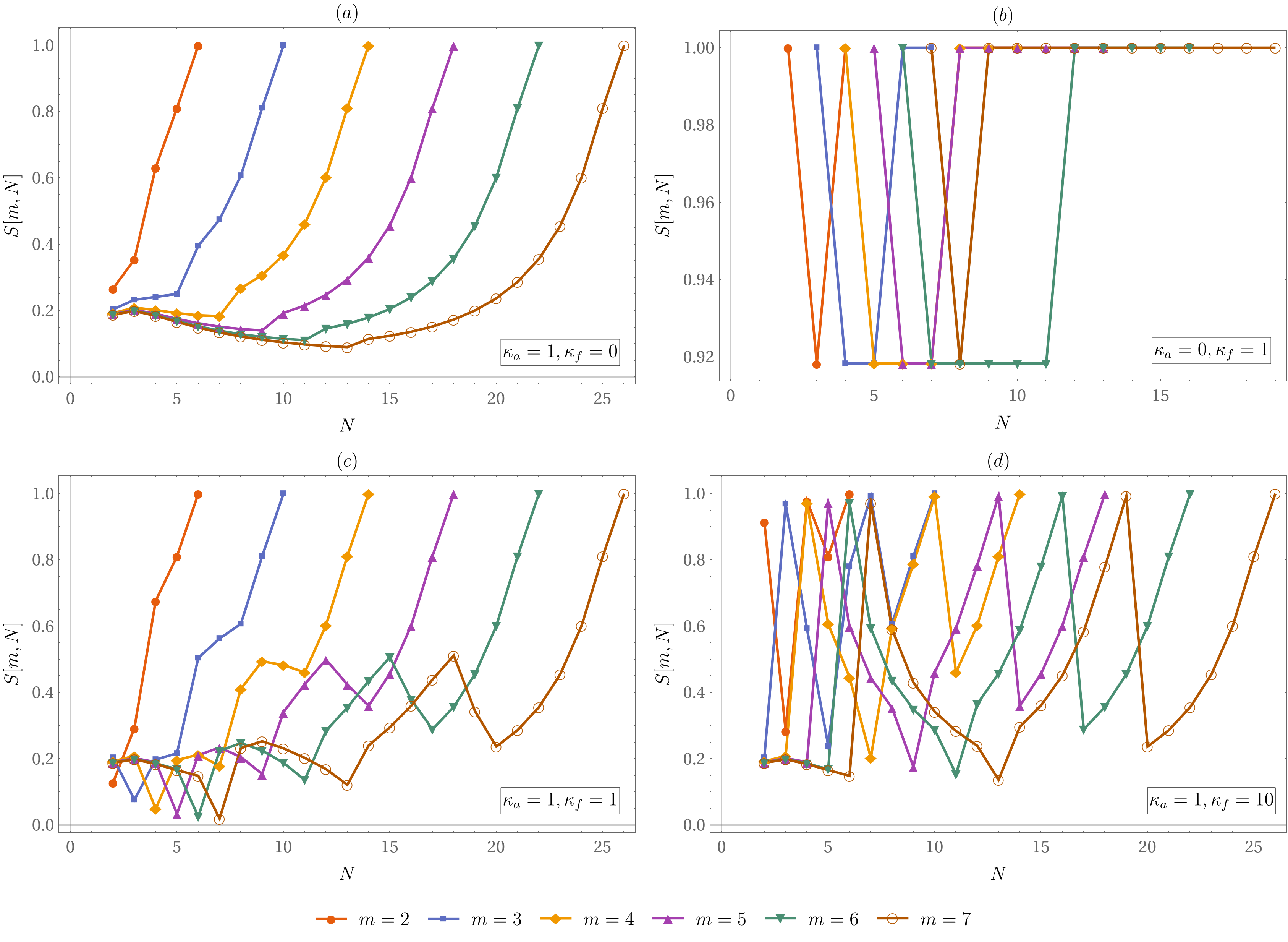}
\caption{Entropy $ S[m,N] $ plotted against the number of particles $ N $ 
for constant values of $ m $ and for different values of $ \kappa_a $ and 
$ \kappa_f $ (the values are given in the inset.)}
\label{entropyplot}
\end{figure}

To find the entanglement entropy, we find the one-particle reduced density 
matrix first. Let $ \ket{\Phi_N^{(0)}} $ be the $ N- $particle ground state. 
We expand the $N-$particle ground state in the $ N- $particle 
Hilbert space basis $  \lbrace \ket{\chi^{(N)}_j} \rbrace  $ where 
\begin{equation}
\ket{\chi^{(N)}_j} = (a^\dagger_1)^j (a^\dagger_2)^{(N-j)} \ket{0}
\end{equation}
and 
\begin{align*}
\begin{aligned}
j& = \begin{cases}
0,1,..,N ~~&\text{if}~~ 0 \leq N\leq 2m-1 \\
N-(2m-1),N-(2m-1)+1,..,(2m-1) ~~&\text{if}~~ 2m-1 < N\leq 2(2m-1) \\
\end{cases}
\end{aligned}
\end{align*}
Thus, 
\begin{align}
\begin{aligned}
\ket{\Phi_N^{(0)}} & = \begin{cases}
 \sum\limits_{j=0}^N \phi^{(N)}_{j} \ket{\chi_j}  ~~&\text{if}~~ 0 \leq 
N\leq 2m-1 \\
\sum\limits_{j=N-(2m-1)}^{2m-1} \phi^{(N)}_{j} \ket{\chi_j}  
~~&\text{if}~~2m-1 < N\leq 2(2m-1) 
\end{cases}
\end{aligned}
\end{align}
Using the following expression
\begin{align}
\begin{aligned}
\bra{0} a_2^{N-1-l} a_1^{l} a_k \ket{\Phi_N^{(0)}} &  =  \delta_{k,1} 
\braket{\chi^{(N)}_{l+1}  |  \Phi_N^{(0)} } +  e^{-i\frac{l}{m^2} } 
\delta_{k,2} \braket{\chi^{(N)}_{l}  |  \Phi_N^{(0)} } \\
& = \delta_{k,1}\phi^{(N)}_{l+1} + e^{-i\frac{l}{m^2}}\delta_{k,2}   
\phi^{(N)}_{l}
\end{aligned}
\end{align}
the reduced density matrix is given by
\begin{align}
\begin{aligned}
\rho^{(N-1)}_{l,l'} & = 
\frac{1}{\mathcal{N}} \rl{  \phi^{(N)}_{l+1}  \rl{\phi^{(N)}_{l'+1}}^* + 
e^{-i\frac{(l-l')}{m^2}}    \phi^{(N)}_{l}    \rl{\phi^{(N)}_{l'}}^*   }  
\end{aligned}
\end{align}
The von Neumann entropy of the reduced density matrix can be found using the 
expression in equation \ref{entropy}. We solve the eigenvalue equation 
\ref{eq:eigenvalue} numerically and obtain the ground state of the 
$N-$particle Hamiltonian for various values of $ N $ and $ m $. One of the 
states is chosen randomly if the ground state is degenerate. The 
corresponding one-particle reduced density matrices and entropy are found 
numerically. In Fig \ref{entropyplot}, we plot the entropy $ S[m,N] $ against 
the number of particles $ N $ for constant $ m $ values for different values 
of $ \kappa_a $ and $ \kappa_f $. 
As expected, the entropy varies with the statistical parameter $ m $ and 
the number of particles $ N $. 

In Fig \ref{entropyplot} a, we plot the entropy corresponding to the anyonic 
hopping model by setting $ \kappa_a=1 $ and $ \kappa_f =0$. In this plot a 
transition point is observed at $ N=2m-1 $. When $ N \leq 2m-1 $, the number 
of configurations (dimension of the $ N $-particle Hilbert space) increases 
linearly with $ N $. But, it decreases linearly with $ N $ for $ N > 2m-1 $ 
since the maximum occupancy at a given lattice site is limited to $ 2m-1 $. 
This transition at $ N=2m-1 $ is reflected in the entropy.

In Fig \ref{entropyplot} b, we plot the entropy corresponding to the fermionic 
hopping model by setting $ \kappa_a=0 $ and $ \kappa_f =1$. For each constant 
value of $ m $, we plot the entropy against $ N $ where $ m \leq N \leq 3m-2$. 
The reason, as mentioned in the last section, is that the $ N $-particle 
Hamiltonian vanishes if $ N<m $ and $ N>3m-2 $ and hence it is meaningless to 
define the entropy in these cases. It is observed that the ground state is 
degenerate in this case and the entropy $ S[m,N] $ for a given value of $ m $ 
and $ N $ depends on the randomly chosen ground state.

In Fig \ref{entropyplot} c and in Fig \ref{entropyplot} d, we plot the entropy 
corresponding to the values $ ( \kappa_a=1,\kappa_f=1 ) $ and $ 
( \kappa_a=1,\kappa_f=10 ).  $ Note that there are several transition points 
at different values of $ N $ for different values of the statistics parameter 
$ m $ in these plots. These transitions mainly occur at values $ N=m $, 
$ N=2m-1 $, and $ N=3m-1 $. The reason for a transition point at $ N=2m-1 $ is 
as explained in the case of Fig. \ref{entropyplot} a. The transition points at 
$ N=m $ and $ N=3m-1 $ are related to the number of fermions in the system. As 
explained before, the entropy does not depend upon the fermionic hopping 
parameter $ \kappa_f $ when $ N<m $ and $ N>3m-2 $ since the term vanishes in 
these limits. The dependence of entropy on the statistics parameter is 
dominated by the anyonic behaviour as the number of particles is increased, 
interspersed with jumps associated with fermionic behaviour obtained when the 
number of anyons completes a fermionic composite.

There are other transitions that happen at different $ N $ values. 
It is not clear if these are an artifact of the numerical calculation, 
or if there is a deeper physical reason.
\section{Conclusions}
Composite particles help us to explain various physical phenomena. In the 
present work, we constructed fermions as composites of anyons in one 
dimension. These anyons are constructed by generalizing Pauli's exclusion 
principle as well as the exchange properties of multiparticle wave functions 
and are defined by a specific algebra we propose. We studied a model of 
these anyons defined by a Hamiltonian which consists of hopping terms both for
elementary anyons and fermionic composites of anyons. We also studied the 
quantum entanglement of these anyons. The dependence of the entanglement 
entropy on the statistics parameter and the number of particles is explored 
numerically.

\appendix

\section{Number operator}\label{numberoperator}
We can derive the following identities using straightforward algebra 
\begin{align}
a_j^p (a_j^\dagger) a_j^q &= a_j^{p+q-1}\\
a_j  (a_j^\dagger)^{p} 
&=  (a_j^\dagger)^{(p-1)}- (a_j^\dagger)^{(\nu)} (a_j)^{(\nu-(p-1))}\\
(a_j)^q  (a_j^\dagger)^{p} &= \delta_{p,q}+ \theta_1(q-p) (a_j)^{q-p}  
+\theta_1(p-q)  (a_j^\dagger)^{(p-q)} - (a_j^\dagger)^{(\nu-(q-1))}
 (a_j)^{(\nu-(p-1))} \\
a_j^{\nu} (a_j^\dagger)^p a_j^{\nu} &=0, ~~~ p<\nu\\
a_j^{\nu} (a_j^\dagger)^{(\nu)} a_j^{\nu} &=a_j^{\nu}
\end{align}
where 
\begin{align}
\begin{aligned}
\theta_1(x) = \begin{cases}
1 & x>0 \\ 0 & x\leq 0
\end{cases}
\end{aligned}
\end{align}
The number density operator is given by
\begin{align}
\begin{aligned}
n_j = \sum_{k=1}^{\nu} (\ad_j)^k a_j^k
\end{aligned}
\end{align}
We compute the commutation relation
\begin{align}
\begin{aligned}
\left[n_j,a_j\right] & =  n_j a_j-a_j n_j\\
& = \sum_{k=1}^{\nu-1} (\ad_j)^k a_j^{k+1} - \sum_{k=1}^{\nu-1} a_j(\ad_j)^k 
a_j^{k} \\
& = \sum_{k=1}^{\nu-2} (\ad_j)^k a_j^{k+1} - 
\sum_{k=1}^{\nu-1} \rl{(a_j^\dagger)^{(k-1)}- (a_j^\dagger)^{(\nu-1)} 
(a_j)^{(\nu-(k-1))}} a_j^{k} \\
& = \sum_{k=1}^{\nu-2} (\ad_j)^k a_j^{k+1} - \sum_{k=1}^{\nu-1} (a_j^\dagger)^{(k-1)} a_j^{k} \\
& = \sum_{k=1}^{\nu-2} (\ad_j)^k a_j^{k+1} - \sum_{k=0}^{\nu-2} 
(a_j^\dagger)^{(k)} a_j^{k+1} \\
& = -a_j
\end{aligned}
\end{align}
Also, 
\begin{align}
\begin{aligned}
\left[n_j,a_k\right] & =0, ~~ k\neq j
\end{aligned}
\end{align}
Therefore we obtain the relations
\begin{align}
\begin{aligned}
\left[n_j,a_k\right] = - \delta_{j,k} a_k,~~~~\left[n_j,a_k^\dagger\right] =  
\delta_{j,k} a_k^\dagger
\end{aligned}
\end{align}
The total number operator
\begin{equation}
\hat{N} = \sum_i{n_i}
\end{equation}
and 
\begin{equation}
\left[\hat{N},a_k\right] = -  a_k,~~~~\left[\hat{N},a_k^\dagger\right] = 
 a_k^\dagger
\end{equation}
\section{Distribution function}\label{app:distri}
We consider a free gas of anyons in the grand canonical ensemble described by the Hamiltonian
\begin{align}
\begin{aligned}
H=\sum_j \epsilon_j n_j
\end{aligned}
\end{align}
where $ n_j $ is the number density operator. The partition function is given by $ Z = \tr (e^{ -\beta (H-\mu N)}) $, where $ \beta $ is the inverse temperature, $ \mu $ is the chemical potential and $ N $ is the total number operator. The number density distribution is given by
\begin{align}
\begin{aligned}
\braket{n_j}& =\frac{1}{Z}\tr(n_j e^{-\beta (H-\mu N)})& =\frac{1}{Z} \sum_{p=1}^{\nu} \tr\rl{(\ad_j)^p a_j^p  e^{-\beta (H-\mu N)}  }
\end{aligned}
\end{align}
We have 
\begin{align}
\begin{aligned}
\left[\hat{N},(a_k^\dagger)^p\right] & =\left[\hat{N},(a_k^\dagger)\right](a_k^\dagger)^{p-1} +(a_k^\dagger)\left[\hat{N},(a_k^\dagger)\right](a_k^\dagger)^{p-2} +..+(a_k^\dagger)^{p-1} \left[\hat{N},(a_k^\dagger)\right] \\&= p (a_k^\dagger)^p.
\end{aligned}
\end{align}
Similarly,
\begin{align}
\begin{aligned}
\left[H,(a_k^\dagger)^p\right] & = p \epsilon_k (a_k^\dagger)^p.
\end{aligned}
\end{align}
Using these relations
\begin{align}
\begin{aligned}
\braket{(\ad_j)^p a_j^p} & = \frac{1}{Z} \tr((\ad_j)^p a_j^p e^{-\beta (H-\mu N)})
\\
& = \frac{1}{Z} \tr\rl{ e^{-\beta (H-\mu N)}  (\ad_j)^p e^{\beta (H-\mu N)} e^{-\beta (H-\mu N)}  a_j^p   }
\\
& = \frac{1}{Z}e^{-\beta p (\epsilon_j-\mu )} \tr\rl{e^{-\beta (H-\mu N)}  a_j^p   (\ad_j)^p }
\\
& = \frac{1}{Z} e^{-\beta p (\epsilon_j-\mu )} \tr\rl{e^{-\beta (H-\mu N)} \rl{ 1- (a_j^\dagger)^{(\nu-(p-1))}(a_j)^{(\nu-(p-1))} }}
\\
& = e^{-\beta p (\epsilon_j-\mu )}  \rl{1- \braket{(a_j^\dagger)^{(\nu-(p-1))}(a_j)^{(\nu-(p-1))}} }.
\end{aligned}
\end{align}
We made use of the relation
\begin{align}
\begin{aligned}
a_j^p (a^\dagger_j)^p & = 1- (a_j^\dagger)^{(\nu-(p-1))}(a_j)^{(\nu-(p-1))} 
\end{aligned}
\end{align}
This gives
\begin{align}
\begin{aligned}
\braket{(\ad_j)^{(\nu-(p-1)) }a_j^{(\nu-(p-1)) }} & = 
 e^{-\beta {(\nu-(p-1)) } (\epsilon_j-\mu )}  \rl{1- \braket{(a_j^\dagger)^{p}(a_j)^{p}} }.
\end{aligned}
\end{align}
The term
\begin{align}
\begin{aligned}
\braket{(\ad_j)^p a_j^p}  & = e^{-\beta p (\epsilon_j-\mu )}  - e^{-\beta p (\epsilon_j-\mu )} \rl{e^{-\beta {(\nu-(p-1)) } (\epsilon_j-\mu )}  - e^{-\beta {(\nu-(p-1)) } (\epsilon_j-\mu )}\braket{(a_j^\dagger)^{p}(a_j)^{p}}}
\\
& = e^{-\beta p (\epsilon_j-\mu )}  - \rl{e^{-\beta {(\nu+1) } (\epsilon_j-\mu )}  - e^{-\beta {(\nu+1) } (\epsilon_j-\mu )}\braket{(a_j^\dagger)^{p}(a_j)^{p}}}.
\end{aligned}
\end{align}
Therefore
\begin{align}
\begin{aligned}
\braket{(a_j^\dagger)^{p}(a_j)^{p}} \rl{1 - e^{-\beta {(\nu+1) }(\epsilon_j-\mu )} }  &=e^{-\beta p (\epsilon_j-\mu )}  - e^{-\beta {(\nu+1) }(\epsilon_j-\mu )} 
 \\
\braket{(a_j^\dagger)^{p}(a_j)^{p}}  &=\frac{e^{-\beta p (\epsilon_j-\mu )}  - e^{-\beta {(\nu+1) }(\epsilon_j-\mu )}  }{\rl{1 - e^{-\beta {(\nu+1) }(\epsilon_j-\mu )} } }
\end{aligned}
\end{align}
Substituting, we get
\begin{align}
\begin{aligned}
\braket{n_j} & = \sum_{p=1}^{\nu} \braket{(\ad_j)^p a_j^p}
\\
 &=\sum_{p=1}^{\nu} \frac{e^{-\beta p (\epsilon_j-\mu )}  - e^{-\beta {(\nu+1) }(\epsilon_j-\mu )}  }{\rl{1 - e^{-\beta {(\nu+1) }(\epsilon_j-\mu )} } }
 \\
&= \frac{ \frac{e^{-\beta  (\epsilon_j-\mu )} - e^{-\beta  (\nu+1)(\epsilon_j-\mu )} }{1- e^{-\beta  (\epsilon_j-\mu )}}  - \nu e^{-\beta {(\nu+1) }(\epsilon_j-\mu )}  }{\rl{1 - e^{-\beta {(\nu+1) }(\epsilon_j-\mu )} } }
 \\
&= \frac{1}{  \rl{e^{\beta  (\epsilon_j-\mu )} -1}} - \frac{ (\nu +1) \rl{e^{\beta (\epsilon_j-\mu )} -1}}{\rl{e^{\beta {(\nu+1) }(\epsilon_j-\mu )} -1}   }
\end{aligned}
\end{align}
\section{Fermionic Algebra}\label{app:fermalg}
We use the notation $ m = \frac{\nu+1}{2}  $ where $ \nu=1,3,5,.. $ and 
$m\in \mathbb{N} $. In this notation, the algebraic relations are
\begin{align}
\begin{aligned}
a_j a_k & =e^{i \theta  \sgn(j-k) } a_k a_j, ~ k\neq j\\
a_j a_k^\dagger & =e^{-i \theta \sgn(j-k) } a_k^\dagger a_j ~ k\neq j\\
a_j a_j^\dagger & = 1-  (a_j^\dagger)^{(2m-1)} (a_j)^{(2m-1)} \\
a_j^{2m}&=0\\
(a_j^\dagger)^{2m}&=0
\end{aligned}
\end{align}
and the identities are
\begin{align}
a_j^p (a_j^\dagger) a_j^q &= a_j^{p+q-1}\\
a_j  (a_j^\dagger)^{p} 
&=  (a_j^\dagger)^{(p-1)}- (a_j^\dagger)^{(2m-1)} (a_j)^{(2m-p)}\\
(a_j)^q  (a_j^\dagger)^{p} &=  \theta_1(q-p) (a_j)^{q-p}  +\theta_1(p-q)  
(a_j^\dagger)^{(p-q)} - (a_j^\dagger)^{(2m-q)} (a_j)^{(2m-p)}\\
a_j^{2m-1} (a_j^\dagger)^p a_j^{2m-1} &=0, ~~~ p<(2m-1)\\
a_j^{2m-1} (a_j^\dagger)^{(2m-1)} a_j^{2m-1} &=a_j^{2m-1}
\end{align}
We define the fermionic operators
\begin{align}
\begin{aligned}
f_j = a_j^{m}
\end{aligned}
\end{align}
Since $  a_j^{2m} =0 $, we have 
\begin{align}
\begin{aligned}
f_j^{2} = 0.
\end{aligned}
\end{align}
Also we derive the commutation relation
\begin{align}
\begin{aligned}
f_j f_k & = a_j^{m} a_k^{m}\\
& = e^{im \theta \sgn(j-k)} a_k  a_j^{m} a_k^{m-1}\\
& = e^{i m^2 \theta\sgn(j-k)} a_k^{m}  a_j^{m}
\end{aligned}
\end{align}
We choose $ \theta = \frac{\pi}{m^2} $. This implies 
\begin{align}
f_j f_k  = -f_k f_j 
\end{align}
Also,
\begin{align}
\begin{aligned}
f_j f^\dagger_k & = a_j^{m}( a_k^\dagger)^{m}\\
& = e^{-i m^2 \theta\sgn(j-k)} ( a_k^\dagger)^{m}  a_j^{m} \\
& = e^{-i \pi\sgn(j-k)} ( a_k^\dagger)^{m}  a_j^{m} \\
& = - f^\dagger_k f_j
\end{aligned}
\end{align}
Similarly, the other relations can be derived
\begin{align}
\begin{aligned}
f_j f^\dagger_j & = a_j^{m}( a_j^\dagger)^{m}\\
&=1-(a_j^\dagger)^{(m)} a_j^{(m)}\\
& = 1-f_j^\dagger f_j
\end{aligned}
\end{align}
Hence, $ f_j $ and $  f_j^\dagger  $ satisfy fermionic algebra.
\section{Matrix representation of the Hamiltonian}\label{app:hammatrix}
 The Hamiltonian for the two-site case is given by
\begin{align}
\begin{aligned}
H&=H_a+H_f\\ H_a& = -\kappa_a \rl{ a_{1}^\dagger a_2+a_{2}^\dagger a_{1}} 
\\H_f &=-\kappa_f \rl{ (a_{1}^\dagger)^m (a_2)^m+(a_{2}^\dagger)^m (a_{1})^m}
\end{aligned}
\end{align}
We choose the following basis for the Fock space
\begin{align}
\begin{aligned}
\lbrace  \ket{r,s} \rbrace ,~~~r ,s =0,1,2,..,\nu
\end{aligned}
\end{align}
where $  \ket{r,s} = (a^\dagger_1)^r (a^\dagger_2)^s\ket{0}  $. 
For general $ \nu $ there will be $ \nu^2 $ basis states when we consider 
two lattice sites.

We can express the Hamiltonian in the matrix form in the 
basis given above
\begin{align}
\begin{aligned}
H_{q,p;r,s} & = \bra{q,p} H \ket{r,s}=\braket{0| a_2^{p} a_1^{q}H  
(a^\dagger_1)^{r} (a^\dagger_2)^{s} |0}
\end{aligned}
\end{align}
We have the following matrix elements
\begin{align}
\begin{aligned}
\bra{q,p} (a^\dagger_1) (a_2) \ket{r,s}&= \bra{0}a_2^p a_1^q  (a^\dagger_1) 
(a_2) (a^\dagger_1)^r (a^\dagger_2)^s\ket{0} 
\\
&=e^{\frac{i \pi }{m^2} } e^{-\frac{i \pi q }{m^2} }\bra{0}a_2^{p+1} a_1^q   
(a^\dagger_1)^{r +1}(a^\dagger_2)^s\ket{0} 
\\
&=e^{\frac{i \pi (1-q)}{m^2} } \delta_{q,r+1} \delta_{p+1,s}
\\
&=e^{-\frac{i \pi r}{m^2} } \delta_{q,r+1} \delta_{p+1,s}
\\
\bra{q,p} (a^\dagger_2) (a_1) \ket{r,s}& =  e^{\frac{i \pi (r-1)}{m^2}  }  \delta_{q+1,r} \delta_{p,s+1} \\
\bra{q,p} (a^\dagger_1)^m (a_2)^m \ket{r,s}& = e^{\frac{i \pi (r-1)}{m^2}  }  \delta_{q+1,r} \delta_{p,s+1}  \\
\bra{q,p} (a^\dagger_2)^m (a_1)^m \ket{r,s}& =  e^{\frac{i \pi m (r-1)}{m^2}  }  \delta_{q+m,r} \delta_{p,s+m} 
\end{aligned}
\end{align}
Using the above results, the matrix elements of the Hamiltonian are
\begin{align}
\begin{aligned}
H_{q,p;r,s} = \bra{q,p} H \ket{r,s}&= -\kappa_a \rl{e^{\frac{i\pi (1-q)}{m^2}}
\delta_{q,r+1} \delta_{p+1,s}+  e^{\frac{i \pi (r-1)}{m^2}} \delta_{q+1,r} 
\delta_{p,s+1} }\\& -\kappa_f \rl{ e^{\frac{i \pi m(1-q)}{m^2} }\delta_{q,r+m}
\delta_{p+m,s}+e^{\frac{i \pi m (r-1)}{m^2}  }  \delta_{q+m,r} \delta_{p,s+m} }
\end{aligned}
\end{align}
\bibliographystyle{unsrt}
\bibliography{global_references.bib}
\end{document}